\documentclass[11pt]{article}
\usepackage{moriond,epsfig}
\usepackage{comment}

\def\Journal#1#2#3#4{{#1} {\bf #2}, #3 (#4)}

\def\PRL{\em Phys. Rev. Lett.}

\def\be{\begin{equation}}
\def\ee{\end{equation}}
\def\bea{\begin{eqnarray}}
\def\eea{\end{eqnarray}}

\def\met{\ensuremath{E\kern-0.6em\lower-.1ex\hbox{/}_T}}
\begin{document}
\vspace*{4cm}
\title{Search for SM Higgs in the $WH \rightarrow l\nu b\bar{b}$ Channel using $\sim$2fb$^{-1}$}

\author{{\bf Tatsuya Masubuchi} on behalf of the CDF Collaboration}

\address{Doctoral Program in Physics, Graduate School of Pure and Applied Sciences, \\
University of Tsukuba, \\Ten'noudai 1-1-1,
 Tsukuba, Ibaraki 305-8571, Japan}

\maketitle

\abstracts{
We report a search for Standard Model (SM) Higgs boson production in association with a $W^{\pm}$ boson. This search uses data corresponding to an integrated luminosity of $\mathrm{1.9\,fb^{-1}}$ collected with the CDF detector at Tevatron. We select events matching the $W$ + jets signature and require at least one jets to be identified as $b$-quark jets. To further increase discrimination between signal and background, we use kinematic information in an artificial neural network.
The number of tagged events and the resulting neural network output distributions are consistent with the Standard Model expectations, and we set an upper limit on the $WH$ production cross section times branching ratio $\sigma(p\bar{p} \rightarrow W{^\pm}H)\times BR(H\rightarrow b\bar{b}) < 1.1$ to $1.0$ pb for Higgs masses from 110 GeV/$c^{2}$ to 150 GeV/c$^2$ at 95\% confidence level.
}
\section{Introduction}
The success of the Standard Model in explaining and predicting experimental data provides strong motivation for the existence of a neutral Higgs boson. Current electroweak fits combined with direct searches from LEP2 indicate the mass of the Higgs boson is less than $190\,\mathrm{GeV}/c^2$ at 95\% confidence level~\cite{lepewwg}.
In proton-antiproton collisions of $\sqrt{s}=1.96\,\mathrm{TeV}$ at the Tevatron, the Standard Model Higgs boson may be produced in association with a $W$ boson~\cite{higgs_xsec}. For low Higgs masses (below $140\,\mathrm{GeV}/c^2$) the dominant decay mode is $H\rightarrow b\bar b$.  The final state from the $WH$ production is therefore $\ell \nu b \bar b$, where the high-$p_T$ lepton from the $W$ decay provides an ideal trigger signature at CDF. The analysis strategies make use of $b$-tagging algorithm to suppress the $W$+jets backgrounds and apply artificial neural network to discriminate signal to remaining backgrounds.
\section{Event Selection}
Events are collected by the CDF II detector with high-$p_T$ central electron or muon triggers which have an 18 GeV threshold.  The central electron or muon is further required to be isolated with $E_T$ (or $p_T$) $>20\,\mathrm{GeV}$ in offline.
Events having the $W$+jets signature are confirmed with a missing transverse energy requirement ($\met > 20\,\mathrm{GeV}$).  

We use forward (plug) electron events with a trigger intended for $W$ candidate events. This trigger requires both a forward electron candidate and missing transverse energy. Plug electron events are further required to have $E_T >$ 20 GeV and $\met > 25$ GeV. For plug electron events, additional selection is required to suppress QCD background.

Events are required to have 2 jets with $E_T > 20\,\mathrm{GeV}$ and $|\eta| < 2.0$. Because the Higgs boson decays to $b \bar{b}$ pairs, we employ $b$-tagging algorithms which relies on the long lifetime and large mass of the $b$ quark to suppress enormous $W$+jets backgrounds.

\subsection{Bottom Quark Tagging Algorithm}
To greatly reduce the backgrounds to this Higgs search, we require that at least one jets in the event be identified as containing $b$-quarks by one of three $b$-tagging algorithms.  The secondary vertex tagging algorithm identifies $b$ quarks by fitting tracks displaced from the primary vertex.  
In addition, we add jet probability tagging algorithm that identifies $b$ quarks by requiring a low probability that all tracks contained in a jet originated from the primary vertex, based on the track impact parameters~\cite{jetprob}.  To be considered for double tag category, an event is required to have either two secondary vertex tags, or one secondary vertex tag and one jet probability tag. 

Furthermore we also make use of exactly one $b$-tagged events with the secondary vertex tagging algorithm. To improve signal-to-background ratio for one tag events, we employ neural network (NN) $b$-tagging algorithm~\cite{Higgs_955} applied. This neural network is tuned for only jets tagged by the secondary vertex tagging algorithm. The purity of $b$-jets tagged by this algorithm is improved. 

Finally we categorize events in three $b$-tagged conditions, double secondary vertex tagged events, one secondary vertex tagged plus one jet probability tagged events and one NN tagged events. These two categories of double-tagged events and one category of one neural network tagged events are defined exclusively.

\subsection{Expected Signal Events and Systematics}
The acceptance in Higgs mass of $120\,\mathrm{GeV}/c^2$ in central region is  0.48$\pm$0.05\%, 0.38$\pm$0.04\%, 0.93$\pm$0.05\% for the double secondary vertex tagged, the secondary vertex plus jet probability and  one neural network tagged category, respectively.
The expected signal events in Higgs mass 120 GeV/c$^{2}$ are about 3.9 events in central plus plug data. 

The uncertainties on the signal acceptance currently have the largest effect on the Higgs sensitivity.  The $b$-tagging uncertainty is dominated by the uncertainty on the data/MC scale factor. 3.5-9.1\% systematic is assigned for each $b$-tagging category. The uncertainties due to initial state radiation and final state radiation (2.9-5.2\%) are estimated as difference from the nominal.  Other uncertainties on parton distribution functions ($\sim$ 2\%), jet energy scale (2-3\%), trigger efficiencies ($< 1\%$) and lepton identification contribute (2\%) are taken into account.  
\section{Backgrounds}
This analysis builds on the method of background estimation detailed in Ref.~\cite{Higgs_955}.  In particular, the contributions from the following individual backgrounds are calculated: falsely $b$-tagged events, $W$ production with heavy flavor quark pairs, QCD events with false $W$ signatures, top quark pair production, and electroweak production (diboson, single top).

We estimate the number of falsely $b$-tagged events (mistags) by counting the number of negatively-tagged events, that is, events in which the measured displacement of the secondary vertex is opposite the $b$ jet direction.  Such negative tags are due to tracking resolution limitations, but they provide a reasonable estimate of the number of false positive tags after a correction for material interactions and long-lived light flavor particles.

The number of events from $W$ + heavy flavor is calculated using information from both data and Monte Carlo programs.  We calculate the fraction of $W$ events with associated heavy flavor production in the ALPGEN Monte Carlo program interfaced with the PYTHIA parton shower code.  This fraction and the $b$-tagging efficiency for such events are applied to the number of events in the original $W$+jets sample after correcting for the $t\bar t$ and electroweak contributions.

We constrain the number of QCD events with false $W$ signatures by assuming the lepton isolation is independent of \met\ and measuring the ratio of isolated to non-isolated leptons in a \met\ sideband region. The result in the tagged sample can be calculated in two ways: by
applying the method directly to the tagged sample, or by estimating the number of non-W QCD events in the pretag sample and applying an average $b$-tagging rate.

In this analysis, 83, 90 and 805 events for each $b$-tagging category are observed against 80.62$\pm$18.75, 86.99$\pm$17.99 and 809.61$\pm$159.38 expected in signal region. The good agreement is also obtained in plug region.

\section{Artificial Neural Network}
\begin{figure}
\begin{center}
\includegraphics[width=5cm,height=4cm]{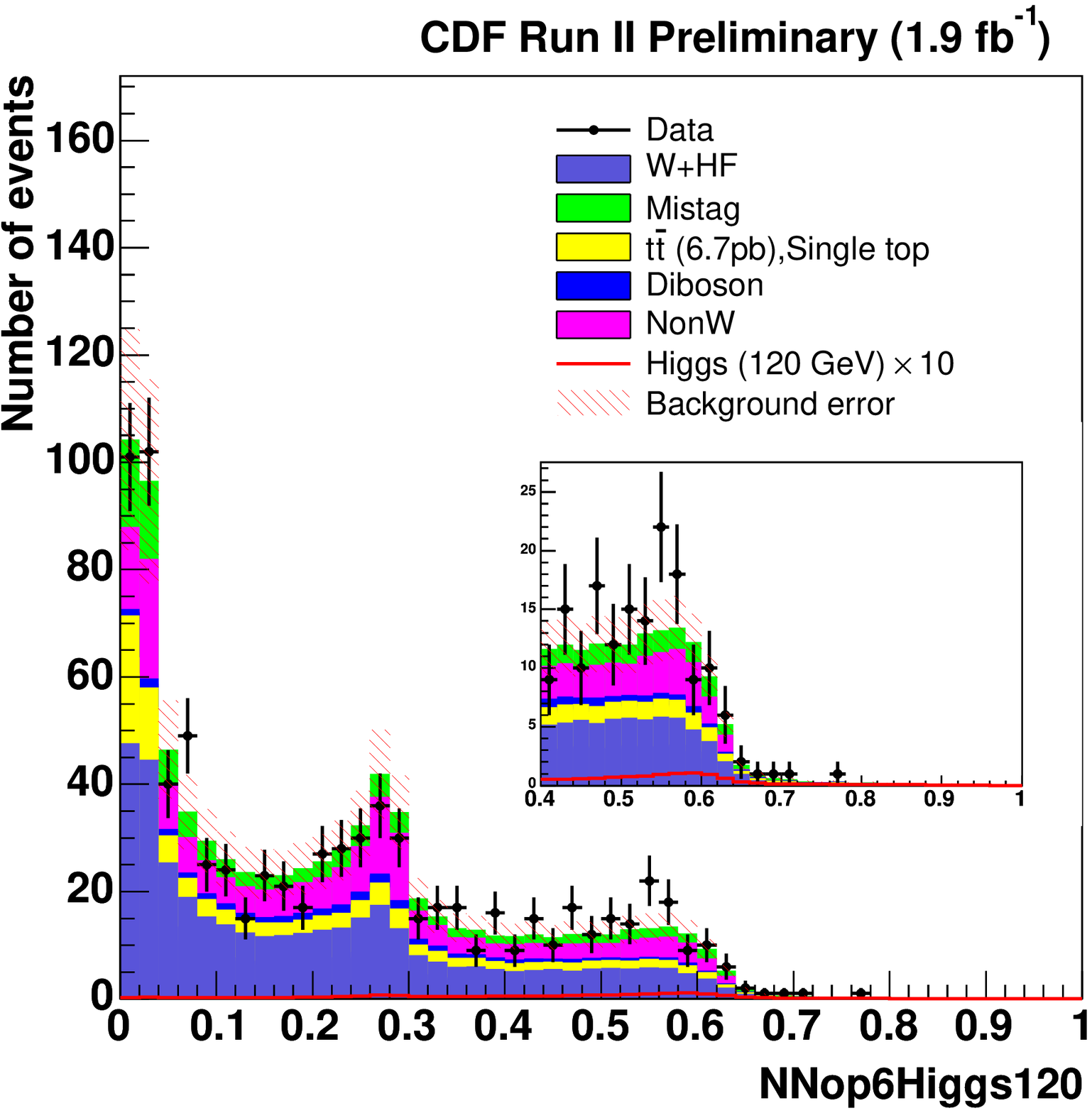}
\includegraphics[width=5cm,height=4cm]{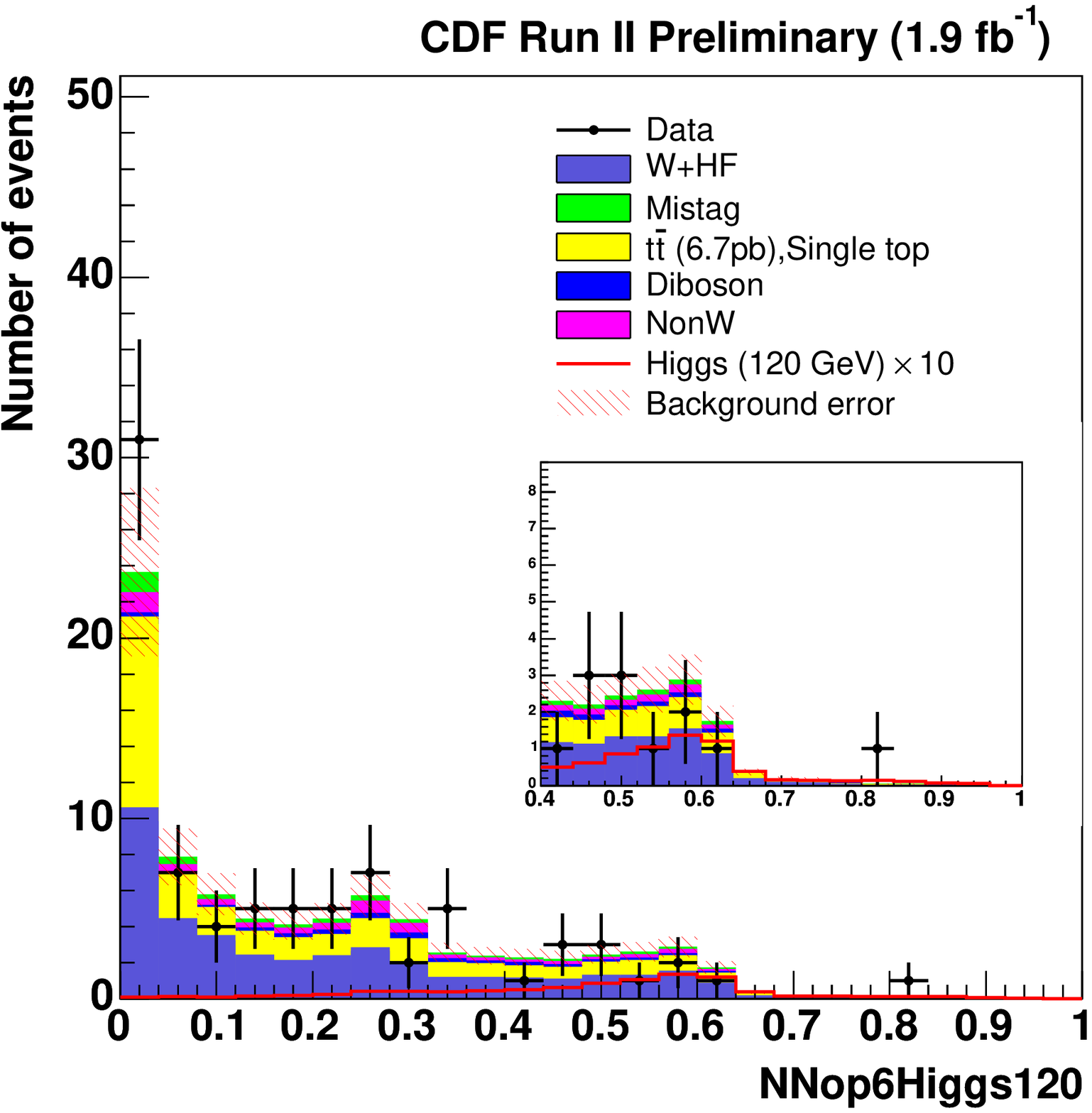}
\includegraphics[width=5cm,height=4cm]{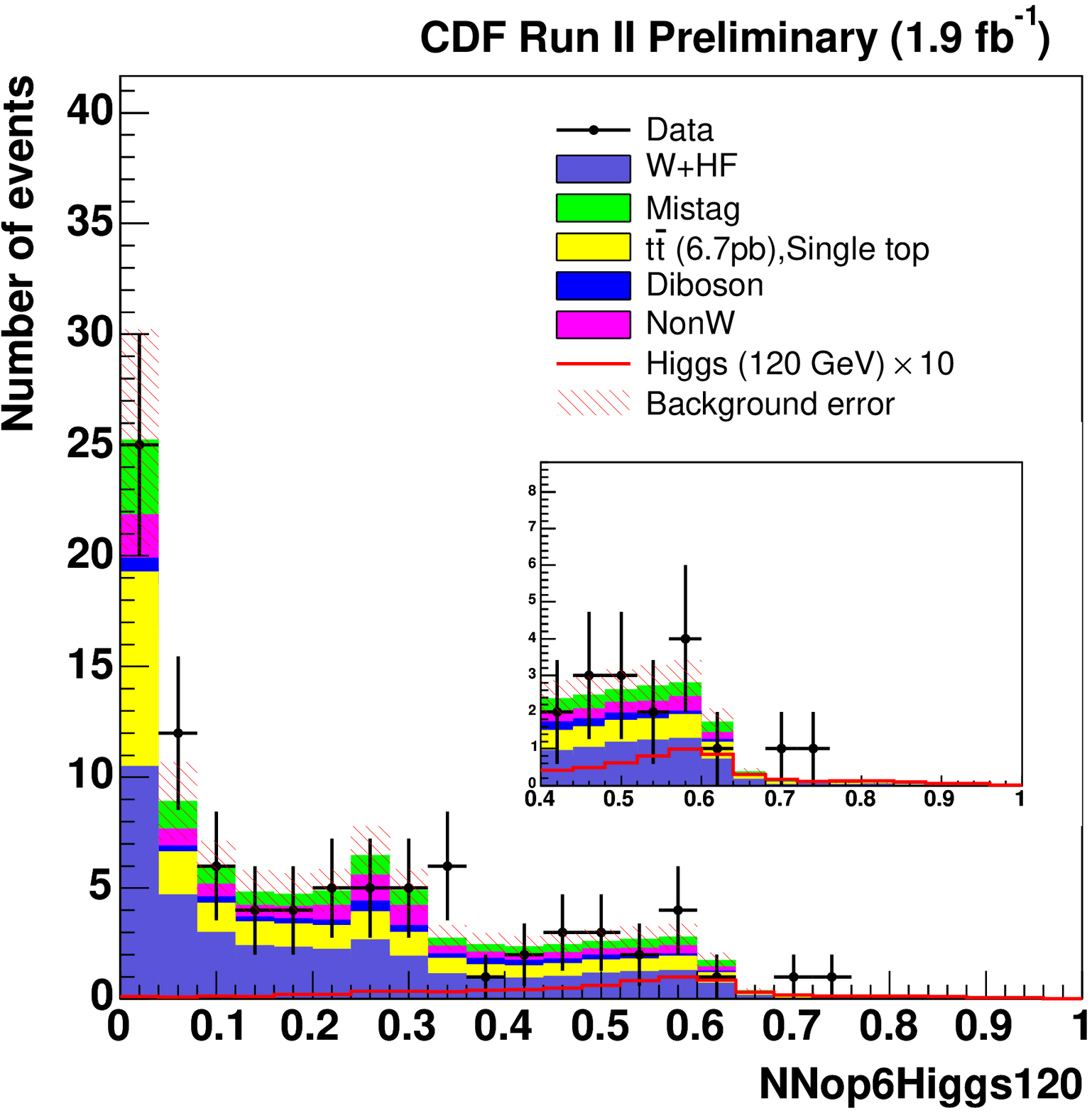}\\
\includegraphics[width=5cm,height=4cm]{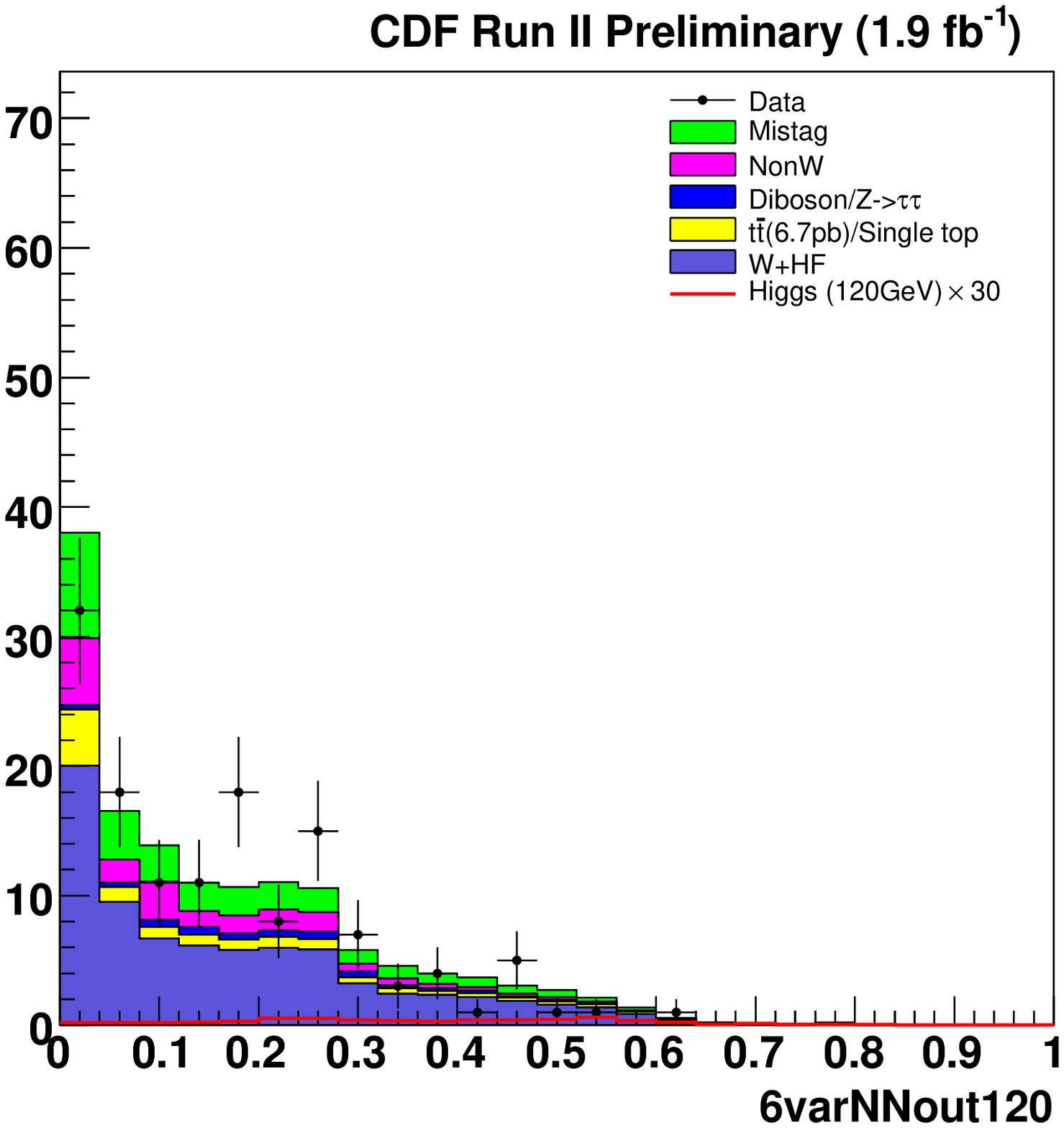}
\includegraphics[width=5cm,height=4cm]{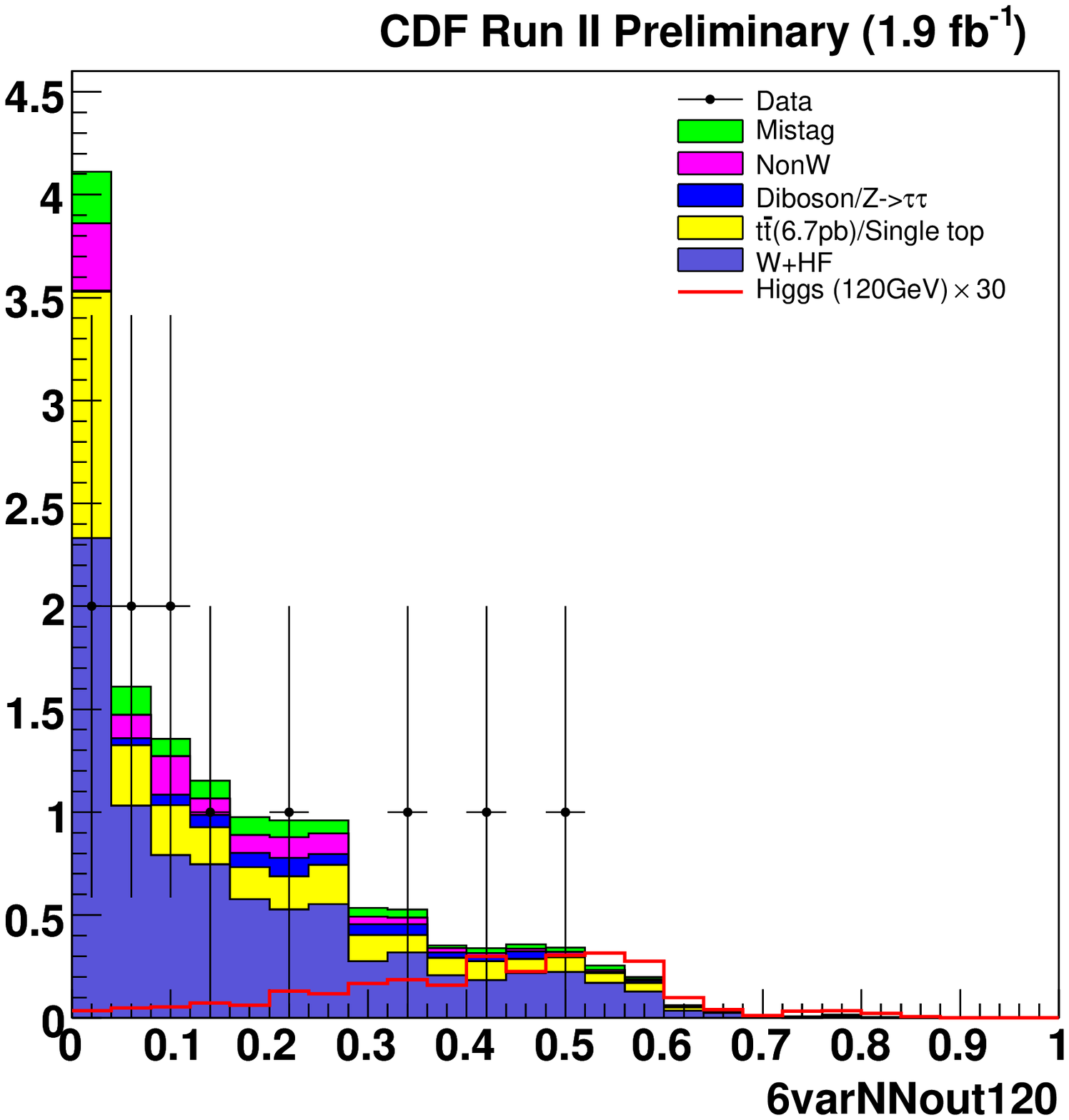}
\includegraphics[width=5cm,height=4cm]{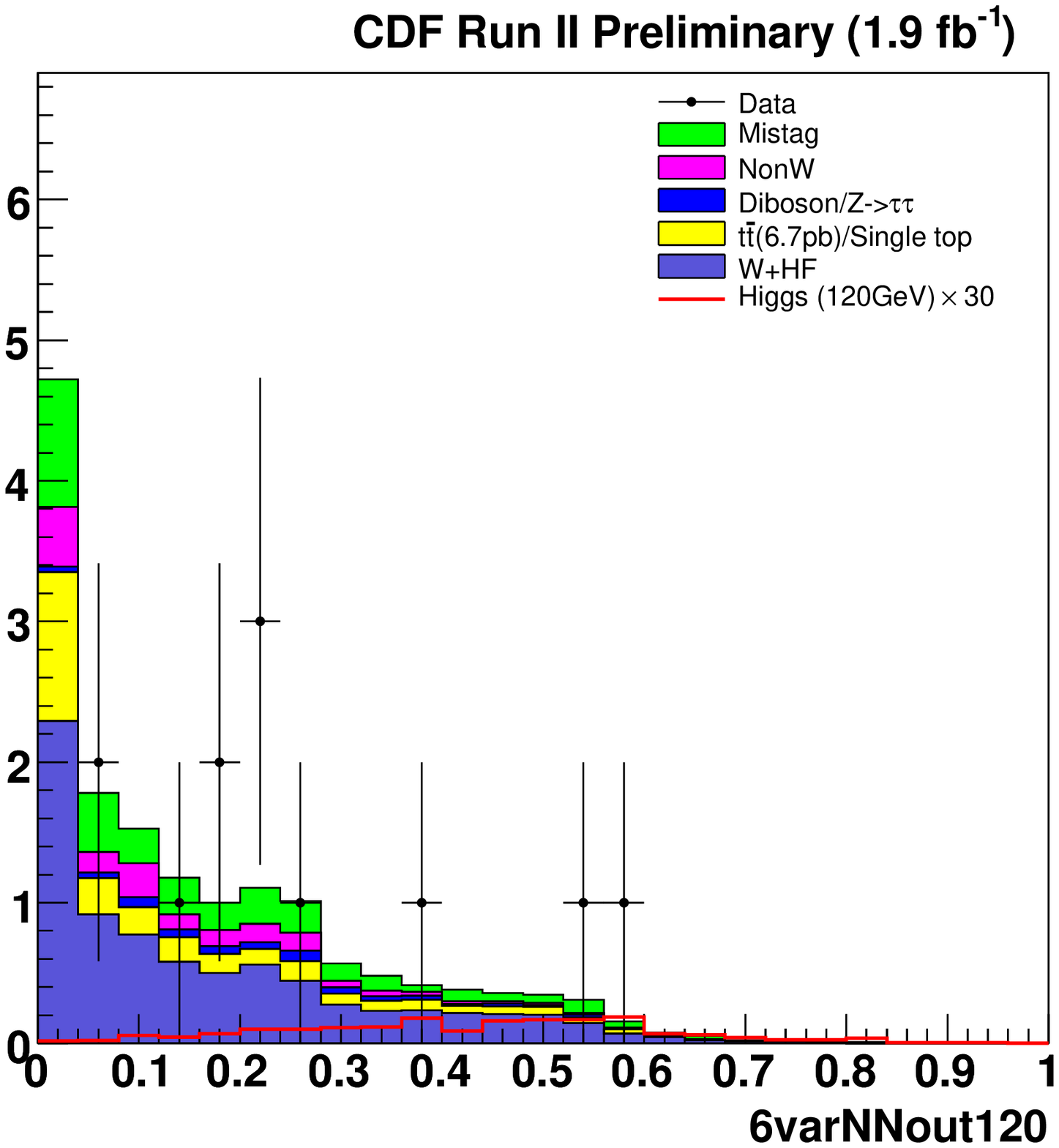}
\caption{Neural network results calculated from six variables for each lepton region (Top:central lepton, Bottom: plug electron). The plot on the left shows exactly one $b$-tagged events, the plots on the center and right show double secondary vertex $b$-tagged event and one secondary vertex plus one jet probability $b$-tagged events, respectively.}
\label{fig:NNcentral}
\end{center}
\end{figure}

To further improve signal to background separation we employ an artificial neural network.  This neural network combines six kinematic variables into a single function with better discrimination between the Higgs signal and the background processes than any of the variables individually.  To train the neural network, JETNET package~\cite{JETNET}.  The input variables are defined below:
\begin{description}
        \item[Dijet invariant mass+:]  The invariant mass reconstructed from the two jets. If there are additional looser jets, the loose jet that is closest to one of the two jets is included in this invariant mass calculation.
\vspace{-0.25cm}
        \item[Total System $p_T$:]  The vector sum of the transverse momenta of the lepton, the \met, and the two jets.
\vspace{-0.25cm}
        \item[$p_T$ Imbalance:]  The scalar sum of the lepton and jet transverse momenta minus the \met. 
\vspace{-0.25cm}
        \item[$\sum E_{T}$ (loose jets):] The scalar sum of the loose jet transverse energy. 
\vspace{-0.25cm}
        \item[$M_{l\nu j}^{min}$:] The invariant mass of the lepton, \met and one of the two jets, where the jet is chosen to give the minimum invariant mass. The $p_{z}$ of neutrino is ignored for this quantity.
\vspace{-0.25cm}
        \item[$\Delta R$ (lepton-$\nu$):] The distance between the direction of lepton and neutrino in $\eta
 - \phi$ plane, where the $p_{z}$ of the neutrino is taken from largest $|p_{z}|$ calculated from $W$ mass constraint.      
\end{description}
The training is defined such that the neural network attempts to produce an output as close to 1.0 as possible for Higgs signal events and as close to 0.0 as possible for background events.  For optimal sensitivity, a different neural network is trained for each Higgs mass considered. Fig.\ref{fig:NNcentral} show the neural network results for each $b$-tagging category. 
\section{Results}
\begin{figure}[h,b]
  \begin{tabular}{cc} 
    \begin{minipage}{0.50\textwidth}
\begin{center}
\includegraphics[width=8.0cm,height=6.0cm]{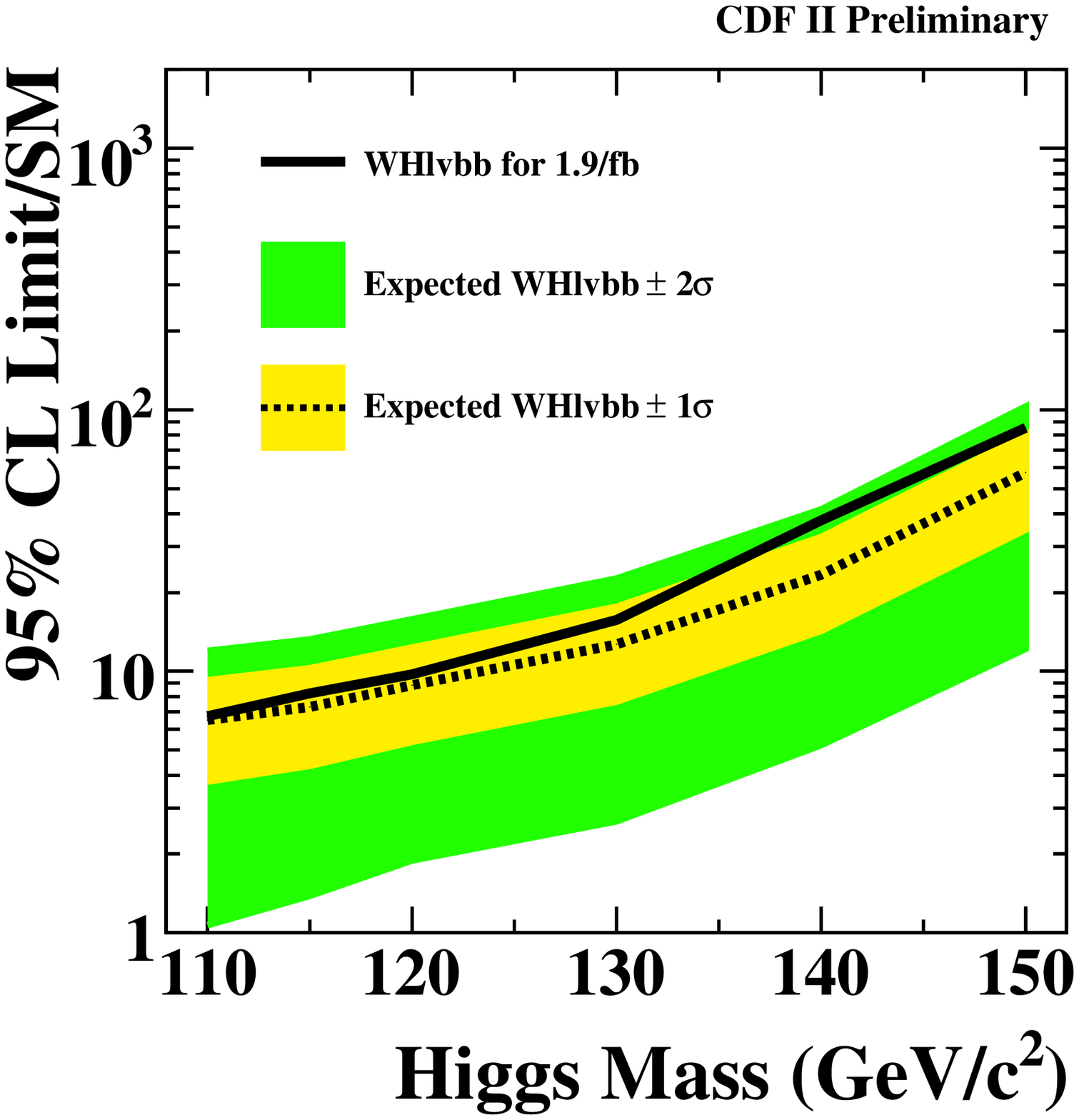}
\caption{Observed and expected limits as a function of the Higgs mass hypothesis. The solid line shows observed 95\% C.L. upper limit and dash line shows expected limit obtained by the assumption of null signal hypothesis.}
\label{fig:Limit}
\end{center}  
    \end{minipage} &
    \begin{minipage}{0.45\textwidth}
\begin{center}
\begin{tabular}{|c|c|c|}\hline
Higgs Mass & \multicolumn{2}{|c|}{\small Upper Limit (pb) / (pb/SM)} \\
(GeV/c$^{2}$) & Observed & Expected \\ \hline
110 & 1.1 (6.8) & 1.1 (6.5) \\ 
115 & 1.1 (8.2) & 1.0 (7.3) \\
120 & 1.1 (9.8) & 1.0 (8.9) \\
130 & 1.0 (15.8) & 1.0 (12.6) \\
140 & 1.2 (37.8) & 0.7 (23.4) \\
150 & 1.0 (85.2) & 0.7 (57.6) \\ \hline
\end{tabular}
\caption{Observed and expected 95\% C.L. upper limit on $\sigma(p\bar{p}\rightarrow WH) \times (\textrm{BR}(H\rightarrow b\bar{b})$. The numbers in parenthesis show the values normalized by the Standard Model theoretical cross section.}
\label{table:limit}
\end{center}
    \end{minipage}
  \end{tabular}
\end{figure}

We perform a direct search for an excess in the signal region of the neural network output distribution from single-tagged and double-tagged $W$+2 jet events. A binned maximum likelihood technique is used to estimate upper limits on Higgs production by constraining the number of backgrounds to the estimates within uncertainties. Each $b$-tagging category is combined to obtain best sensitivity for both central and forward lepton region. We set an upper limit on the production cross section times branching ratio as a function of $m_H$, plotted in Fig.~\ref{fig:Limit}.  The results are also collected in Table~\ref{table:limit}.

\end{document}